\documentclass{PoS}

\usepackage{amsmath,amsfonts,amssymb,xspace}
\usepackage[arrow,curve,matrix,arc,rotate]{xy}

\newcommand{\IZ}{\mathbb{Z}}
\newcommand{\kk}{k}

\newcommand{\IR}{\mathbb{R}}
\newcommand{\IC}{\mathbb{C}}
\newcommand{\IN}{\mathbb{N}}

\newcommand{\IP}{\mathbb{P}}

\newcommand{\pa}{\partial}

\newcommand{\be}{\begin{equation}}
\newcommand{\ee}{\end{equation}}
\newcommand{\ben}{\begin{eqnarray}\displaystyle}
\newcommand{\een}{\end{eqnarray}}

\newcommand\bOm{\bar\Omega}

\newcommand\Om{\Omega}

\newcommand{\cA}{\mathcal{A}}
\newcommand{\cB}{\mathcal{B}}
\newcommand{\cH}{\mathcal{H}}
\newcommand{\cL}{\mathcal{L}}
\newcommand{\cM}{\mathcal{M}}
\newcommand{\cN}{\mathcal{N}}
\newcommand{\cW}{\mathcal{W}}
\newcommand{\cX}{\mathcal{X}}
\def\Tr{\,{\rm Tr}\, }

\newcommand{\dx}{c}
\newcommand{\de}{{\rm d}}
\newcommand{\I}{{\rm i}}

\newcommand{\gref}{g_{\rm Coulomb}}
\newcommand{\gR}{G_{\rm Higgs}}

\newcommand{\OmS}{\Omega_{\rm S}}  
\newcommand{\bOmS}{\bOm_{\rm S}}  

\newcommand{\empha}[1]{#1}
\newcommand{\emphb}[1]{#1}
\newcommand{\emphc}[1]{#1}

\newcommand{\ttz}{t}

\title{Corfu lectures on wall-crossing, multi-centered black holes, and quiver invariants}

\ShortTitle{Wall-crossing, multi-centered black holes, and quiver invariants}

\author{\speaker{Boris Pioline}$^{ab}$,\\
\llap{$^a$} CERN Dep PH-TH, 1211 Geneva 23, Switzerland \\
\llap{$^b$} Laboratoire de Physique Th\'eorique et Hautes Energies, CNRS UMR 7589,   \\
Universit\'e Pierre et Marie Curie - Paris 6, 4 place Jussieu,
75252 Paris cedex 05, France\\
E-mail: \email{boris.pioline@cern.ch}}

\abstract{The BPS state spectrum  in four-dimensional gauge theories or string vacua with $\cN=2$ supersymmetries is well known to depend on the values of the parameters or moduli at spatial infinity. The
BPS index is locally constant,  but  discontinuous across real codimension-one walls where
some of the BPS states decay. By postulating that BPS states are bound states of more elementary 
constituents carrying their own degrees of freedom and interacting via supersymmetric quantum mechanics, we provide a physically transparent derivation of the universal wall-crossing formula which governs the jump of the index. The same physical picture suggests that at any point in moduli space, the total index can be written as a sum of contributions from all possible bound states of elementary, absolutely stable constituents with the same total charge. For D-brane bound states described by quivers, this `Coulomb branch formula' predicts that the cohomology of quiver moduli spaces is
uniquely determined by certain `pure-Higgs' invariants, which are the microscopic analogues of single-centered
black holes. These lectures are based on joint work with J. Manschot and A. Sen. 
}

\FullConference{Proceedings of the Corfu Summer Institute 2012 "School and Workshops on Elementary Particle Physics and Gravity",
		September 8-27, 2012, 
		Corfu, Greece}

\begin{document}

\section{Introduction}

Determining the spectrum of BPS states in four-dimensional gauge theories and string vacua with 
$\cN=2$ supersymmetries has been an active topic of research since the seminal work of Seiberg and Witten \cite{Seiberg:1994rs}. Indeed, it affords a window into strongly coupled phenomena triggered by non-perturbative states, such as monopole confinement and condensation in softly broken $N=2$ theories or conifold transitions between apparently disconnected Calabi-Yau (CY) vacua, and allows for precision tests of non-perturbative dualities. Notably, it remains a challenge to determine the exact degeneracies
of BPS black holes in $\cN=2$ string vacua, and study deviations from the classical
Bekenstein-Hawking entropy at finite electromagnetic charges, to the same level of precision as has been
achieved for $\cN=4$ or $\cN=8$ string vacua (see e.g. \cite{Denef:2007vg,Sen:2007qy,Banerjee:2011jp,Dabholkar:2012nd,Sen:2011ba} for the current state of the art).

While the BPS spectrum is largely insensitive to variations of parameters or coupling constants, as BPS states lie in short multiplets and can only desaturate the BPS bound by combining with another BPS state of opposite parity,  the BPS index $\Omega(\gamma;t)=\Tr (-1)^{2J_3}$ is in general {\it not } 
constant over the space $\cB$ of parameters $t$: the reason is that along certain real-codimension
one walls $\cW$ in $\cB$, known as walls of marginal stability, the point-particle spectrum meets with the continuum of multi-particle states, signaling  that a BPS state with charge $\gamma$ is about to decay
 into more elementary BPS constituents with charge $\alpha_i$ with the same total charge $
\gamma=\sum_i \alpha_i$ \cite{Seiberg:1994rs}\footnote{A similar phenomenon was first encountered for  two-dimensional  theories with $(2,2)$ supersymmetry in \cite{Cecotti:1992rm}.}. 
Across the wall, the total BPS index $\Omega(\gamma;t)$ jumps by a quantity which is equal to the BPS index of the bound state, which contributes on one side of $\cW$ but
not on the other. While this wall-crossing phenomenon complicates the analysis of the BPS spectrum, it turns out to be  governed by a completely universal formula, which holds both in gauge theories and string 
vacua. This formula first appeared in various guises in the mathematics literature on Donaldson-Thomas (DT) invariants \cite{ks,Joyce:2008pc}, and has received various physics proofs since then \cite{Denef:2007vg,Diaconescu:2007bf,Gaiotto:2008cd,Chuang:2010wx,Gaiotto:2010be,Andriyash:2010qv,Manschot:2010qz,Stoppa:2011}.

In the first lecture, we shall outline a physically transparent derivation of this wall-crossing formula \cite{Manschot:2010qz}, which hinges on a general solution of the supersymmetric quantum mechanics of multi-centered solitons. While being equivalent to the original wall-crossing formulae of \cite{ks,Joyce:2008pc}, 
its structure is completely different 
and has already led to a number of new mathematical insights on 
DT invariants \cite{ReinekeMPS,Mozgovoy:2012} (see \cite{Pioline:2011gf} 
for an earlier review of wall-crossing, with a
different emphasis).

\begin{figure}
\centerline{\includegraphics[height=4cm]{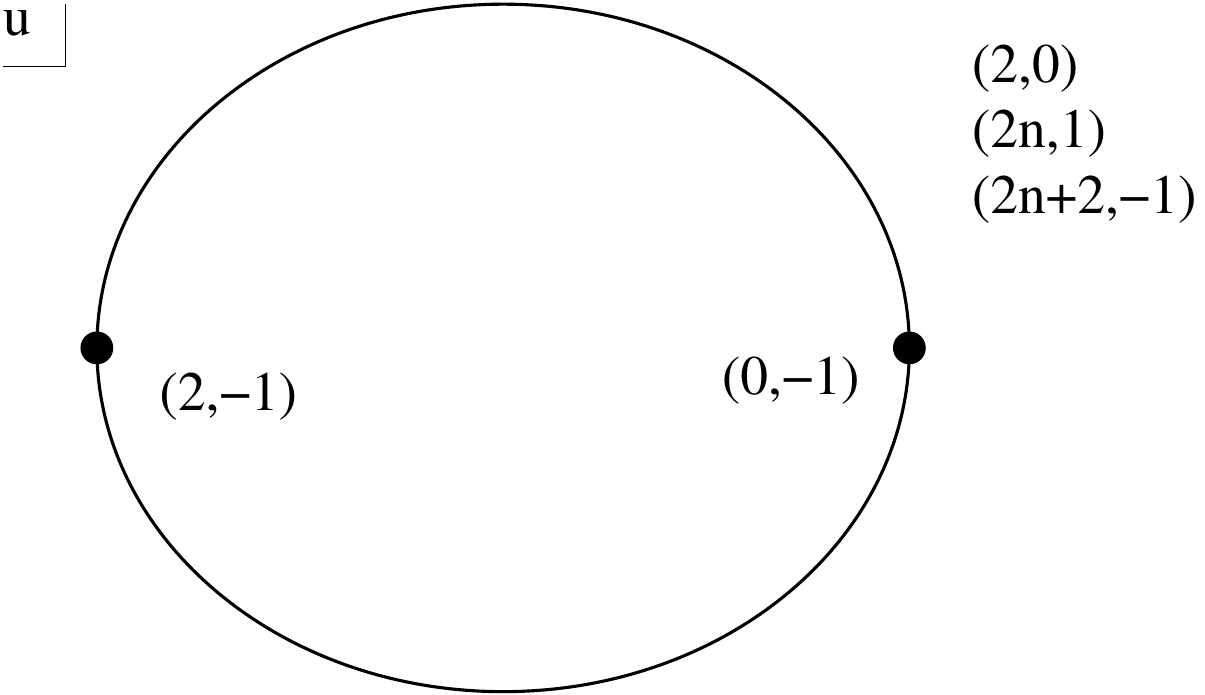} }
\caption{ BPS spectrum 
in $\cN=2, D=4$ SYM theory with $SU(2)$  gauge group and no flavor. 
In the strong coupling chamber, the only
stable BPS states in the strong coupling chamber are the monopole
and dyon, with charge $(0,-1)$ and $(2,-1)$. All states in the weak coupling chamber outside the wall
arise as bound states of the monopole and dyon.  
\label{fig_sw}}
\end{figure}

One implication of wall-crossing is that at least some of the BPS states are not elementary, being liable to decay into more elementary constituents. There are reasons to believe, however, that there might 
exist a subset of BPS states which are absolutely stable throughout the parameter space $\cB$.
This is certainly the case in the simplest $\cN=2$ gauge theory considered in \cite{Seiberg:1994rs,Bilal:1996sk}, where
the monopole and dyon are the only BPS states in the strong coupling chamber, while all BPS states
in the weak coupling chamber arise as bound states of these two (see Figure \ref{fig_sw}). It is also clear
in string theory vacua that BPS states represented by single-centered black holes cannot decay (except perhaps at the boundary between two basins of attraction \cite{Denef:2001xn}). Thus, it should be possible
to reconstruct the total BPS index $\Omega(\gamma;t)$ from the BPS degeneracies $\OmS(\gamma)$ associated to these `single-centered' or `elementary' constituents, by solving the quantum mechanics
of a BPS bound state of $n$ single-centered constituents with charges $\{\alpha_i\}$. Since the charges are unrestricted, the quantum mechanics is in general more involved that in the case relevant for wall-crossing, as the phase space of multi-centered black holes
can have non-compact regions where some of the centers can approach each other to arbitrarily small distance. 
In Lecture 2
I shall present a general prescription for computing the index of such bound states, which is based on 
the physical hypothesis that these `scaling solutions', which classically have zero angular momentum, carry the smallest possible angular momentum at the quantum level \cite{Manschot:2011xc}.

While it is at present impractical to compute the total BPS index $\Omega(\gamma;t)$ or single-centered BPS index $\OmS(\gamma)$ in general Calabi-Yau string vacua, as it would require determining the complete set of DT invariants of the CY threefold $\cX$, it is sometimes possible to realize a class of BPS states as bound states of certain D-branes wrapped on supersymmetric cycles in $\cX$, and described by some matrix quantum mechanics described
by a certain quiver \cite{Douglas:1996sw,Denef:2002ru}. Mathematically, the BPS bound states are represented by cohomology classes on the moduli space of semi-stable quiver representations. This provide a concrete framework to test
the afore-mentioned hypothesis, and identify the elementary constituents as 
special cohomology classes. 
A recent study of three-node quivers with a loop \cite{Bena:2012hf} and similar necklace quivers \cite{Lee:2012sc}
suggest that these elementary states can be understood as middle cohomology states, or `pure Higgs' states, which turn out to be remarkably robust under wall-crossing. In Lecture 3 we shall review the techniques used to
analyze these quivers and use our general prescription to identify the `pure-Higgs'
states in a more precise way  \cite{Manschot:2012rx}.  

\section{Wall-crossing from multi-center black hole quantum mechanics}

In this first lecture, we discuss wall-crossing phenomena in four-dimensional string vacua described by $\cN=2$ supergravity, or in $\cN=2$ gauge theories described by an Abelian  gauge theory on the Coulomb branch. In both cases, the jump of the index across the wall is universally determined by the quantum mechanics of multi-centered
solitons, which is identical for black holes and dyons.

\subsection{Generalities}

Let $\Gamma=\Gamma_e\oplus \Gamma_m$
be the lattice of electric and magnetic charges $(q_\Lambda,p^\Lambda)$, 
with symplectic (Dirac-Schwinger-
Zwanziger) integer pairing 
\be
\langle \gamma, \gamma' \rangle 
= q_\Lambda p'^{\Lambda} - q'_\Lambda p_\Lambda \in \IZ
\ee
BPS states preserve 4 out of 8 supercharges, and saturate the bound 
\emphb{$M\geq |Z_\gamma(t)|$} where the central charge 
$Z_\gamma(t) 
= \langle Y(t), \gamma \rangle$ is a linear functional of the electromagnetic charges \cite{Witten:1978mh}.
The index 
\be
\label{defOmega}
\Omega(\gamma ; t) = -\frac12 \Tr_{\cH_\gamma(t)} (-1)^{2J_3} (2J_3)^2 = \Tr_{\cH'_\gamma(t)} (-1)^{2J_3}
\ee
(where $\cH_\gamma(t)$ is the Hilbert space of one-particle
states with charge $\gamma\in \Gamma$ in the vacuum with vector moduli $t\in\cB$, and the prime denotes the removal
of the center of motion degrees of freedom)
receives contributions from short multiplets only. 
 In $\cN=2$ gauge theories (but not in $\cN=2$ string vacua), there is an additional $SU(2)_R$
symmetry, which allows to defined  the protected spin character (PSC)
$\Omega(\gamma;t,y)= \Tr (-1)^{2J_3}\, y^{2(I_3+J_3)}$ \cite{Gaiotto:2010be}. 
The `non exotics'  conjecture  \cite{Gaiotto:2010be} 
asserts that all states have $I_3=0$, and the PSC 
coincides with the refined index, defined by 
\be
\label{defOmegay}
\Omega(\gamma;t,y)= \Tr_{\cH'_\gamma(t)} (-1)^{2J_3}\, y^{2J_3}\ .
\ee
In $\cN=2$ SUGRA/string vacua, one can still define $\Omega(\gamma;t,y)$ by \eqref{defOmegay} but it is no longer protected, hence it could get contributions from non-BPS states and depend on HM moduli. Nevertheless, it will be a very useful device to compute the protected index \eqref{defOmega}, obtained by setting $y=1$ in
\eqref{defOmegay}.

The BPS index $\Omega(\gamma;t)$ is locally constant, but it may jump on codimension-one loci 
\be
W(\gamma_1,\gamma_2) = \{ t \in \cB\ : \  Z_{\gamma_1}(t)/Z_{\gamma_2}(t)\in \IR^+ \} 
\ee
in vector-multiplet moduli space, called  `walls of marginal stability'. The basic mechanism is that some of the BPS states with charge $\gamma=M\gamma_1+N\gamma_2$ are bound states of more elementary BPS states with charge $\alpha_i=M_i\gamma_1+N_i\gamma_2$ with $\sum_i(M_i,N_i)=(M,N)$, which exist only on one side of the wall $W(\gamma_1,\gamma_2) $. Several remarks are in order:

\begin{itemize}
\item[i)] for a given wall, we can always choose the basis vectors $\gamma_1,\gamma_2$ to be primitive (i.e.
such that $\gamma_i/d \notin \Gamma$ for all integers $d>1$),
and such that 
$\Omega(M\gamma_1+N\gamma_2)$ has support on the positive cone $\tilde\Gamma=\IN \gamma_1+\IN \gamma_2$ (corresponding
to particles) and its opposite $-\tilde\Gamma$ (corresponding to anti-particules) \cite{Andriyash:2010qv}. 
The constituents lie in $\tilde\Gamma$, so only a finite number of bound states can occur.

\item[ii)]  The index of states with $\gamma\notin \IZ\gamma_1+\IZ \gamma_2$
is constant across the wall $W(\gamma_1,\gamma_2)$, and so are $\Omega(\gamma_1)$ and $\Omega(\gamma_2)$.

\item[iii)]  The index $\Omega(M\gamma_1+N\gamma_2)$ may contain contributions from bound states of constituents with charges lying outside the sublattice $\IZ\gamma_1+\IZ \gamma_2\subset\Gamma$, but those are insensitive to the wall. 

\item[iv)]  The states $\{\alpha_i\}$ do not form a BPS bound state on the wrong side of the wall, but one cannot
exclude the possibility that they could form a non-BPS bound state. Such states do not contribute to $\Delta\Omega(\gamma)$, nor to the PSC, but in string vacua they could contribute to the refined index $\Delta \Omega(\gamma;y)$. We shall work under the assumption that this possibility does not occur.

\end{itemize}

\subsection{Primitive wall-crossing}

As a warm-up we consider the fate of BPS states with charge $\gamma=\gamma_1+\gamma_2$ across
the wall $W(\gamma_1,\gamma_2)$. For $\langle \gamma_1,\gamma_2\rangle\neq 0$, there exists a two-centered BPS solution of charge $\gamma=\gamma_1+\gamma_2$, angular momentum
$\vec J=\frac12 \langle \gamma_1, \gamma_2\rangle \, \vec r/|\vec r|$ of the low energy effective action, 
with separation given by \cite{Denef:2000nb}
\be
|\vec r| = \frac
{\langle \gamma_1,\gamma_2\rangle\, | Z(\gamma_1+\gamma_2)|}{2\, \Im[\bar Z(\gamma_1)\, Z(\gamma_2)]}
\ee
The solution exists only on one  side of the wall. 
As $t$ approaches the wall, the distance $|\vec r|$ diverges and the bound state 
decays into its constituents $\gamma_1$ and $\gamma_2$. 
Near the wall, the two centers can be treated as \empha{pointlike charged particles} with  
$\Omega(\gamma_i)$ internal degrees of freedom, interacting 
via exchange of massless particles, including the usual Coulomb and Lorentz forces. 
 The \emphb{classical BPS phase space} $\cM_2$ for the two-particle system is the
two-sphere, with symplectic form \cite{Denef:2002ru}
\be
\omega=
\tfrac12\gamma_{12} \sin\theta\, \de\theta\de\phi\ ,\quad
\gamma_{12}\equiv \langle \gamma_1,\gamma_2\rangle
\ee
such that rotations $\pa_{\phi}$ are generated by the Hamiltonian\footnote{Recall that the Hamiltonian
(or moment map) of a vector field $\kappa$ preserving the symplectic form $\omega$ is the function 
$\mu$ such that $\iota_\kappa \omega = \de \mu$. The additive ambiguity
is uniquely fixed by the condition $\{\mu_{\kappa},\mu_{\kappa'}\}=\mu_{[\kappa,\kappa']}$ in
the case of a   simple Lie group action.}
$J_3=\frac12 \gamma_{12} \cos\theta$.
The geometric quantization of $\cM_2$ leads to  $|\gamma_{12}|$ states transforming as  a spin 
\emphc{$j=\tfrac12(|\gamma_{12}|-1)$} multiplet under rotations. 
Near the wall, these configurational degrees of freedom decouple from the 
 internal degrees of freedom of the two particles. The index of the two-particle bound state is then
$ (-1)^{\gamma_{12}+1} \gamma_{12}   \times\Omega(\gamma_1)\times
\Omega(\gamma_2)$.
These are the only bound states of charge $\gamma=\gamma_1+\gamma_2$ which (dis)appear across the wall, thus the jump of the index is given by the primitive wall-crossing formula \cite{Denef:2007vg}
\be
\label{primcross}
\Delta\Omega( \gamma_1+\gamma_2)= (-1)^{\gamma_{12}+1}  \, \gamma_{12}\, 
\Omega(\gamma_1)\, \Omega(\gamma_2)\ .
\ee
Similarly, the variation of the refined index is given by \cite{Diaconescu:2007bf,Dimofte:2009bv}
\be
\label{primref}
\Delta\Omega( \gamma_1+\gamma_2;y)=
\frac{ (-y)^{\gamma_{12}} - (-y)^{-\gamma_{12}}}{y-1/y}  \,
\Omega(\gamma_1;y)\, \Omega(\gamma_2;y)\ ,
\ee
which reduces to the previous formula at $y=1$. 

\subsection{The quantum mechanics of multi-centered black holes \label{sec_qm}}

We shall now generalize this reasoning to compute $\Delta\Omega(M\gamma_1+N\gamma_2)$
for general $(M,N)$. For this we need a description of the BPS phase space of multi-centered
black holes (or dyons). The most general stationary, $n$-centered BPS solution of $\cN=2$ 
SUGRA with total charge $\gamma=\alpha_1+\cdots+\alpha_n$ is \cite{Denef:2000nb,Bates:2003vx}
\be
\de s^2 = - 
e^{2U} \, (\de t+\cA )^2 + e^{-2U} \de \vec r^2  
\ee
where  the scale factor $U(\vec r)$ and the scalar fields $t(\vec r)$ are determined by 
solving
\be
\label{eqatt}
2\, e^{-U(\vec r)} \Im\left[ e^{-\I\phi} Y\left(t(\vec r)
\right) \right]=  \beta +
\sum_{i=1}^n {\alpha_i\over |\vec r - \vec r_i|} \ .
\ee
The one-form $\cA$ and the electromagnetic fields are uniquely determined, provided the locations $\vec r_i$
satisfy the conditions
\be \label{denef3d}
\empha{\sum_{j=1\atop j\ne i}^n \frac{\langle \alpha_i ,\alpha_j \rangle}{|\vec r_i -\vec r_j|} =  c_i}\ .
\ee
In these equations, we have denoted
\be
\phi=\arg Z_\gamma\ ,\quad \ ,\quad \beta = 2 \, {\rm Im} \left[e^{-\I\phi}\, Y(\ttz_\infty) \right] \ ,\quad
c_i \equiv 2 \, {\rm Im}\, (e^{-\I\phi} Z_{\alpha_i}) \ .
\ee
Eq. \eqref{denef3d} enforces $n-1$ conditions on $3n$ locations $\vec r_i$. Modding out by translations in $\IR^3$, the space of solutions is a $2n-2$ dimensional space which we denote by $\cM_n(\{\alpha_i\}, \{c_i\})$.
For the solution to be regular, one must also ensure 
\be
\label{esk2aa}
e^{-2U(\vec r)}  = \frac{1}{\pi} S\left( \beta +\sum_{i=1}^n {\alpha_i\over |\vec r - \vec r_i|}\right)
> 0\, , \qquad \forall \ \vec r\in \IR^3\, 
\ ,
\ee
where $S(\beta)$  is the Bekenstein-Hawking entropy functional \cite{Bates:2003vx}.
While this condition may in general remove some connected components in $\cM_n$,
it appears to be automatically satisfied when all charges $\alpha_i$ lie in a two-dimensional 
lattice $M\gamma_1+N\gamma_2$ with $MN\geq 0$, the case relevant for wall-crossing.
In this case, the constants $c_i$ are uniquely determined by 
$c_i=\Lambda\sum_{j\neq i}\langle \alpha_i ,\alpha_j \rangle$,
up to an overall scale $\Lambda$ which vanishes on the wall. For $\cN=2$ gauge theories, the most
general multi-centered solution of the low energy Abelian gauge theory on the Coulomb branch has not been worked out, but it is widely believed that solutions are similar to the one above, with the same equilibrium
condition \eqref{denef3d} on the center (but without the constraint \eqref{esk2aa}) \cite{Kim:2011sc}.

When the centers are well-separated, 
the dynamics of the bound state is described by $\cN=4$ quantum mechanics with $3n$ bosonic 
coordinates, $4n$ fermionic coordinates with Lagrangian \cite{Denef:2002ru,deBoer:2008zn,Lee:2011ph,Kim:2011sc}
\be
\cL = \sum_i [W_i(\vec r_i)]^2  + \sum_i \vec A_i \dot {\vec {r_i}} 
+ \sum_{i,j}  G_{ij} \dot {\vec {r_i}} \dot {\vec {r_j}} + \dots
\ee
\be
W_i = \sum_{j\ne i} \frac{\langle \alpha_i ,\alpha_j \rangle}{|\vec r_{ij}|} - c_i\ ,\qquad
\vec A_i = \sum_{j\neq i} \langle \alpha_i, \alpha_j \rangle\, \vec A_{\rm Dirac}(
\vec r_{ij})\ ,
\ee
where $\vec r_{ij}=\vec r_i-\vec r_j$, and $\vec A_{\rm Dirac}(\vec r)$ is the gauge potential of a unit charge Dirac magnetic monopole sitting at $\vec r=0$.
Factoring out the center of mass motion, the classical ground state dynamics is then first order quantum mechanics on the \empha{BPS phase space} $\cM_n(\{\alpha_i\}, \{c_i\}) = \{ W_i=0 \}/\IR^3$, equipped with a symplectic form $\omega$ uniquely determined by the requirement that the moment map of the action of $SO(3)$ rotations  on $\cM_n$ coincides with the total angular momentum $\vec J$ carried by the configuration \cite{Denef:2002ru,deBoer:2008zn}
\be
\omega = \tfrac12 \sum_{i<j} \langle \alpha_i, \alpha_j \rangle\,\de \vec A_{\rm Dirac}(
\vec r_{ij})\ ,\quad
\vec J=\tfrac12 \sum_{i<j}  \langle \alpha_i, \alpha_j \rangle \frac{\vec r_{ij}}{r_{ij}}\ .
\ee
The geometric quantization of $\cM_n$ produces a BPS Hilbert space given by zero-modes
of the natural Dirac operator on $(\cM_n,\omega)$. The refined index (or PSC in gauge theory setting) 
is equal to the \emphb{equivariant Dirac index} \cite{Manschot:2011xc,Lee:2011ph,Kim:2011sc}
\be
\label{equivi}
\gref (\{\alpha_i\}, \{c_i\};y) = \Tr_{{\rm Ker} D_+} (-y)^{2J_3} -  \Tr_{{\rm Ker} D_-} (-y)^{2J_3}\ . 
\ee
For reasons which will become clear in Lecture 3, we refer to \eqref{equivi} as the Coulomb index
of the $n$-centered configuration.
In the limit where all DSZ products $\alpha_{ij}\equiv \langle \alpha_i, \alpha_j \rangle$ are scaled to infinity, the Coulomb index reduces 
to the equivariant \emphb{symplectic volume}
\be
g_{\rm class}(\{\alpha_i\}, \{c_i\};y) =\frac{(-1)^{\sum_{i<j} \langle \alpha_i, \alpha_j \rangle-n+1}}{(2\pi)^{n-1}(n-1)!}
\int_{\cM_n}\, \omega^{n-1}\, y^{2J_3}\ .
\ee
As we shall see, the index can be computed exactly for any number of centers using localization
with respect to rotations $J_3$ around a fixed axis.

\subsection{Localization and the Coulomb index}

Away from walls of marginal stability, defined by hyperplanes $\sum_{i\in A} c_i=0$ in the space 
of the parameters $c_i$ (where $A$ is any proper subset of $\{1,\dots,n\}$), the relative distances 
$|\vec r_{ij}|$ are bounded both from above and from below, $0<r<|\vec r_{ij}|<R$ so that the BPS phase space
$\cM_n$ is compact. Moreover, the action of $J_3$  has only isolated fixed points, namely 
collinear configurations along the $z$ axis,  subject to
the one-dimensional
reduction of \eqref{denef3d},
\be \label{esa1}
\sum_{j\ne i}^n { \langle \alpha_i, \alpha_j \rangle\over |z_{i}-z_j|}
= c_i\, ,\quad
J_3= \frac12 \sum_{i<j}  \langle \alpha_i, \alpha_j \rangle\, {\rm sign}(z_j-z_i)
\, \ .
\ee
These are $n-1$ conditions on $n-1$ locations $z_i-z_1$, hence typically admit isolated solutions.
Thus, both $g_{\rm class}$ and $\gref$ can be evaluated by localization with respect to the
rotations $J_3$ along the $z$ axis, using the Duistermatt-Heckman and  Atiyah-Bott-Lefschetz
fixed point formulae, respectively \cite{Manschot:2010qz,Manschot:2011xc}. For \eqref{equivi} this gives 
\be
\label{AtiyahBott}
\gref (\{\alpha_i\},\{c_i\};y) = 
 \sum_{\rm fixed\  pts} 
\frac{y^{2J_3}}{\det( (-y)^{L} - (-y)^{-L})}
\ee
where $L$ is the matrix of the action of $J_3$ on the holomorphic tangent space 
around the fixed point. With some work, the determinant around a fixed point \eqref{esa1} evaluates
to  $(y-1/y)^{n-1}$ times a sign 
\emphc{$s(p)=-{\rm sign}(\det W'')$} where $W''$ is the Hessian of the superpotential 
\be
\label{defhatW}
W( \{z_i\}) = -\sum_{i<j}  \langle \alpha_i, \alpha_j \rangle \, {\rm sign}[z_j-z_i]\, 
 \ln| z_j - z_i| - \sum_i c_i  z_i \ ,
\ee
whose critical points reproduce the constraints \eqref{esa1}.
After the dust settles, one finds that the refined Coulomb index is given 
by \cite{Manschot:2010qz,Manschot:2011xc}
\be
 \label{elocal55}
\gref(\{\alpha_i\}, \{c_i\}; y)
= \frac{(-1)^{\sum_{i<j} \alpha_{ij} +n-1}}
{(y - y^{-1})^{n-1}} \sum_{p} s(p)\, 
y^{\sum_{i<j}  \langle \alpha_i, \alpha_j \rangle\,  
{\rm sign}( z_j- z_i)}\, 
\ee
where the sum runs over all solutions of \eqref{esa1}.
The contribution of each fixed point is singular at $y=1$, but the sum over
fixed points is guaranteed to produce a symmetric polynomial in $y$ and $1/y$, 
as long as $\cM_n$ is \emphb{compact}. In the simplest two-centered case with
 $c_1\langle \alpha_1, \alpha_2 \rangle>0$, the
fixed points on the BPS phase  $\cM_2=S^2$, lie at the north and south pole,  with $2J_3=\pm \langle \alpha_1, \alpha_2 \rangle$, 
hence \be
\label{gref2}
\begin{split}
\gref(\{\alpha_1,\alpha_2\})=& 
(-1)^{\alpha_{12}+1}  \frac{y^{\langle \alpha_1, \alpha_2 \rangle} -  y^{-\langle \alpha_1, \alpha_2 \rangle}}{y-1/y} 
 = \Tr_{j=\tfrac12(\langle \alpha_1, \alpha_2 \rangle-1)}\,  y^{2J_3} \ .
\end{split}
 \ee
If the sign of $c_1$ is opposite, the BPS phase space is empty and $\gref(\{\alpha_1,\alpha_2\})$ vanishes. Setting $\alpha_1=\gamma_1,\alpha_2=\gamma_2$ and multiplying
by the degeneracies $\Omega(\gamma_i;y)$ carried by each center, one recovers \eqref{primref}.

For more than 2 centers, the enumeration of solutions to \eqref{esa1} becomes rapidly impractical, even
on a computer. It is worth however noting that the contribution of a given ordering of the centers
to \eqref{elocal55} does not depend on the details of the solution, only on the Morse index of the
critical point of $W$, and so is invariant under deformations of the DSZ products $\alpha_{ij}$ and 
constants $c_i$, away from certain singular loci. This robustness was exploited recently to give an
inductive algorithm for computing the Coulomb index, which is much more efficient than looking for 
numerical solutions to \eqref{esa1} \cite{Manschot:2013sya}. 
In the case relevant for wall-crossing, this algorithm gives a 
fully explicit result,
\be \label{eindexfin}
\begin{split}
 \gref(\{\alpha_1,\cdots \alpha_n\}; \{\dx_1,\cdots \dx_n\};y) 
=& (-1)^{n-1+\sum_{i<j} \alpha_{ij}} (y-y^{-1})^{-n+1} \\
 \sum_\sigma 
\prod_{k=1}^{n-1}  \Theta\left(\alpha_{\sigma(k), \sigma(k+1)} \, \sum_{i=1}^k c_{\sigma(k)} \right) &
(-1)^{\sum_{k=1}^{n-1}
\Theta(-\alpha_{\sigma(k),\sigma(k+1)})}\, 
y^{\sum_{i<j} 
\alpha_{\sigma(i)\sigma(j)}} \, ,
\end{split}
\ee
where $\Theta(x)$ is the Heaviside function, equal to $1$ if $x>0$ and $0$ otherwise,
$\sigma$ runs over all permutations of $\{1,\dots,n\}$, and we denote
$\alpha_{ij}=\langle \alpha_i, \alpha_j \rangle$.

\subsection{Statistics and the Coulomb branch wall-crossing formula}

Having computed the index of the quantum mechanics of $n$ centers, we can apply the same logic as before and tentatively write the jump in the total index as a sum of the indices of all bound states that decay across the wall, 
\be
\label{jswcf1}
\Delta\Om(\gamma;y) \stackrel{?}{=}
\sum_{n\geq 2 }\, 
  \sum_{\substack{ \{\alpha_1,\dots \alpha_n\} \in \tilde\Gamma\\
\gamma= \alpha_1+\dots +\alpha_n}
}\, 
g(\{\alpha_i\};y) \prod\nolimits_{i=1}^n \Om^+(\alpha_i;y) 
 \ee
 where $\gamma=M\gamma_1+N\gamma_2,
\alpha_i=M_i\gamma_1+N_i\gamma_2$, and 
$\Omega^+(\alpha_i;y)$ is the refined index on the side where the bound
 state does not exist. This is almost right, but it overlooks the issue of statistics.
If the centers were classical, distinguishable objects, Maxwell--Boltzmann statistics would require to weight
each (unordered) contribution $\{\alpha_i\}$ with a symmetry factor $1/|{\rm Aut}(\{\alpha_i\})|$, 
where $|{\rm Aut}(\{\alpha_i\})|$ is the order of the subgroup of the permutation group leaving $\{\alpha_i\}$
invariant. Instead, the centers are quantum objects, with Bose statistics if $\Omega(\alpha_i)>0$, or 
Fermi statistics if $\Omega(\alpha_i)<0$. One can show that the
Maxwell-Boltzmann prescription nevertheless works, provided one replaces everywhere
the index $\Omega(\gamma)$ with its rational counterpart \cite{Manschot:2010qz}
\be
\label{defbOm}
\bar \Omega(\gamma;y) 
\equiv \sum_{d|\gamma} 
\frac{(y - y^{-1})}{d (y^d - y^{-d})}
\Omega(\gamma/d, y^d) \ ,\quad 
\bar \Omega(\gamma) 
\equiv \sum_{d|\gamma} \frac{1}{d^2} 
\Omega(\gamma/d) \ .
\ee
To see why this can work, consider for example $\Delta\Omega(\gamma_1+2\gamma_2)$: it receives contributions from 
bound states $\{\gamma_1+\gamma_2,\gamma_2\}$, $\{\gamma_1,2\gamma_2\}$,
$\{\gamma_1,\gamma_2,\gamma_2\}$.  The last, 3-centered case can be treated as a two-body problem,
since the two centers with charge $\gamma_2$ have zero DSZ product hence do not interact.  
Taking into account Bose-Fermi statistics for $\{\gamma_1,\gamma_2,\gamma_2\}$, one finds, setting $y=1$
for simplicity,
\be
\begin{split}
\Delta\Om(\gamma_1+2\gamma_2)=& (-1)^{{\gamma_{12}}} {\gamma_{12}}   \Om^+(\gamma_2)\, \Om^+(\gamma_1+\gamma_2) 
+ 2 {\gamma_{12}}  \, \Om^+(2\gamma_2)  \, \Om^+(\gamma_1) \\&
+ \empha{\tfrac12 {\gamma_{12}} \, \Om^+(\gamma_2) \left( \gamma_{12} \Om^+(\gamma_2) +1 \right) }
 \emphb{\Om^+(\gamma_1)}  \ . 
\end{split}
\ee
The last contribution violates the naive charge conservation implied by \eqref{jswcf1}, but is forced
by Bose--Fermi statistics.  Rewriting this formula in terms of the rational invariant $\bOm(2\gamma_2)=\Omega(2\gamma_2)+\tfrac14 \Omega(\gamma_2)$,  one finds that naive charge conservation is restored, 
with the expected $\tfrac12$ Boltzmann symmetry factor,
\be
\begin{split}
\Delta\bOm(\gamma_1+2\gamma_2)=& (-1)^{{\gamma_{12}}} {\gamma_{12}}   \bOm^+(\gamma_2)\, \bOm^+(\gamma_1+\gamma_2) 
+ 2 {\gamma_{12}}  \, \bOm^+(2\gamma_2)  \, \bOm^+(\gamma_1) \\&
+ \empha{\tfrac12 [\gamma_{12} \, \bOm^+(\gamma_2)]^2}
 \emphb{\bOm^+(\gamma_1)}  \ . 
\end{split}
\ee
A general justification for the replacement $\Omega\to\bOm$ can be found in  \cite{Manschot:2010qz}.

After these amendments, we arrive  at the \empha{Coulomb branch wall-crossing formula},
\be
\label{jswcf2}
\Delta\Om(\gamma;y)= \sum_{n\geq 2 }\, 
  \sum_{\substack{ \{\alpha_1,\dots \alpha_n\} \in \tilde\Gamma\\
\gamma= \alpha_1+\dots +\alpha_n}
}\, 
\frac{g(\{\alpha_i\};y)}{|{\rm Aut}(\{\alpha_i\})|} \prod\nolimits_{i=1}^n \bOm^+(\alpha_i;y) 
 \ee
Remarkably, this formula can be shown to agree with the wall-crossing formulae
established in the mathematical literature on Donaldson-Thomas invariants for the
derived category of coherent sheaves on a Calabi-Yau threefold. The agreement is far from
trivial and can be established by induction on the number of centers \cite{Sen:2011aa}. 
As a simple example, the formula \eqref{jswcf2} 
with $\langle \gamma_1,\gamma_2\rangle=2$ 
reproduces the weak coupling spectrum  of  pure $SU(2)$ gauge theory, from the knowledge of the strong coupling spectrum (see Figure \ref{fig_sw}) \cite{Gaiotto:2008cd}.

\section{The total index from single-centered degeneracies}

The wall-crossing phenomenon discussed in Lecture 1 shows that the BPS states contributing to the total refined index (or PSC) $\Omega(\gamma;t;y)$ are in general not elementary, but can be decomposed into more elementary constituents. This suggests the possibility that there could exist a set of  elementary,  absolutely stable BPS states, with moduli-independent indices  $\OmS(\gamma;y)$, such that any other BPS state would arise as a bound state of those. This is indeed realized in pure $SU(2)$ gauge theory, where all states arise as bound states of the monopole and dyon. More generally, it happens in any $\cN=2$
gauge theory which has a strong coupling chamber with a finite number of BPS states. In $\cN=2$ string vacua,
it is clear that BPS states represented by single centered black holes cannot decay and so should be considered as elementary. In this lecture, I shall address the question of computing the total index $\Omega(\gamma;t;y)$ (not only its variation) in terms of indices $\OmS(\gamma;y)$ associated to elementary constituents, without committing to the nature of these constituents.

According to the physical picture explained in Lecture 1, it is natural to expect that the total index 
$\Omega(\gamma;t;y)$, at any point in moduli space, should be a sum of contributions of all possible
bound states of elementary constituents with charge $\alpha_i$ such that $\sum \alpha_i=\gamma$.
Assuming that the BPS Hilbert space associated to each bound state factorizes into configurational
degrees of freedom, described by the geometric quantization of the BPS phase space 
$\cM_n(\{\alpha_i\}, \{c_i\})$, and internal degrees of freedom $\OmS(\alpha_i;y)$ associated to each center,
and taking into account Bose-Fermi statistics,  one may tentatively write 
\be
\label{tentativeC}
\begin{split}
\bOm(\gamma;t;y) = &\sum_{n\geq 1}
\sum_{\substack{ \{\alpha_1,\dots \alpha_n\} \in \Gamma\\
\gamma= \alpha_1+\dots +\alpha_n}}
\frac{\gref(\{\alpha_i\},\{c_i\},y)}{|{\rm Aut}(\{\alpha_i\})|} \, 
\bOm_{\rm S}(\alpha_i,y)\ .
 \end{split} 
\ee
where the rational invariants $\bOm_{\rm S}(\alpha_i,y)$ are defined in terms of the integer invariants $\Om_{\rm S}(\alpha_i;y)$ as in \eqref{defbOm}. 
While similar in spirit with the Coulomb branch wall-crossing formula \eqref{jswcf2}, this formula
differs in several important ways:
\begin{itemize}
\item[i)] it includes a contribution from single centered black holes, with $n=1$;
\item[ii)] the charges $\alpha_i$ run over {\it all} possible vectors in $\Gamma$ with $\bOmS(\alpha_i)\neq 0$, rather 
than a two-dimensional sublattice $\tilde\Gamma$;
\item[iii)] the Coulomb index $\gref(\{\alpha_i\},\{c_i\},y)$ depends on the moduli at infinity $t$ through
the constants $c_i$; in the context of $\cN=2$ supergravity it should only count states consistent with 
the regularity condition \eqref{esk2aa}.  
\end{itemize}
The fact that the charges of the constituents are no longer restricted to the positive cone in a two-dimensional sublattice  has drastic consequences, as we now discuss.

\subsection{Scaling solutions \label{sec_scal}}
Unlike the case relevant for wall-crossing,  the BPS phase space $\cM_n(\{\alpha_i\}, \{c_i\})$ for generic
choices of the DSZ product is in general non-compact, even away from walls of marginal stability.  To see this, consider the three-centered case with DSZ products 
$\alpha_{12}=a$, $\alpha_{23}=b$, $\alpha_{31}=c$ satisfying  triangular inequalities $0<a<b+c$, etc.
It is easy to check that the equations \eqref{denef3d}, 
\be
\frac{a}{r_{12}} - \frac{c}{r_{13}} = c_1, \quad 
\frac{b}{r_{23}} - \frac{a}{r_{12}} = c_2\ ,\quad
\frac{c}{r_{13}} - \frac{b}{r_{23}} = c_3\ ,
\ee
admit solutions where the three centers can approach each other arbitrarily, 
$r_{12}\sim a \epsilon, r_{23}\sim b \epsilon, r_{13}\sim c \epsilon$ with $\epsilon\to 0$ \cite{Denef:2007vg,Bena:2007qc}.
These configurations, known as `scaling solutions', exist for any choice of $c_1,c_2,c_3$ such that $c_1+c_2+c_3=0$, and carry arbitrarily small angular momentum $\vec J^2\sim \epsilon^2$.
For $c_1,c_2>0$, a naive application of the Atiyah-Bott fixed point formula \eqref{elocal55}  gives
\be
\label{gref3}
\gref
=  \frac{(-1)^{a+b+c} (y^{a+b-c}+y^{-a-b+c}) }{(y-1/y)^2}\ ,
\ee
where the two terms arise from the regular collinear configurations $(123)$ and $(321)$. This is {\it not}
a polynomial in $y$, in particular it is singular as $y\to 1$. An explicit computation shows that the equivariant volume of $\cM_3$ is nevertheless finite, and agrees with the Duistermat-Heckman formula after including
by hand an additional fixed point contribution from the scaling boundary of $\cM_3$, with effective angular momentum $J_3=0$ \cite{Manschot:2011xc}. Similarly, \eqref{gref3} can be turned into a Laurent polynomial for any $a,b,c$ by adding by hand a fixed point contribution with $J_3=0$ (when $a+b+c$ is even) or $J_3=\tfrac12$ (when $a+b+c$ is odd):
\be
\tilde g_{\rm Coulomb}
=  \frac{(-1)^{a+b+c}  \left(y^{a+b-c}+y^{-a-b+c}-
\left[ \begin{array}{ccl} 2 &,& a+b+c \ {\rm even} \\
y+1/y &,&  a+b+c \ {\rm odd} 
\end{array} \right]  \right) }{(y-1/y)^2}
\ee
More generally, scaling regions in $\cM_n(\{\alpha_i\},\{c_i\})$  arise whenever there exist
a subset $A$ and vectors $\vec r_i\in \IR^3$, $i\in A$ such that 
\be
\forall i\in A\ ,\quad \sum_{j\in A} \frac{\alpha_{ij}}{|\vec r_{ij}|} = 0\ .
\ee
Such solutions exist, when they do, in all chambers in the space of the parameters $c_i$'s, and have vanishing
angular momentum $\vec J^2$ as the centers approach each other. 
Despite these non-compact regions, one can argue that the symplectic volume is finite, and it is plausible
that $\cM_n$ admits a compactification such that the Atiyah-Bott Lefschetz formula still applies (after
allowing for non-isolated fixed points at the boundary). Indeed, it is possible to formulate a `minimal modification
hypothesis' which converts the Coulomb index $\gref$ computed from \eqref{elocal55} into a symmetric Laurent polynomial $\tilde g_{\rm Coulomb}$, by adding by hand suitable fixed point contributions with the minimum possible angular momentum $J_3$ \cite{Manschot:2011xc}. Unfortunately, this is not good enough to ensure
that the answer from \eqref{tentativeC} will be a symmetric Laurent polynomial, as we now discuss.

\subsection{The Coulomb branch formula for the total index}

Even if the Coulomb indices $\gref(\{\alpha_i\},\{c_i\},y)$ were Laurent polynomials, the tentative formula
\eqref{tentativeC} would not necessarily  produce a  Laurent polynomial for $ \Omega(\gamma;t;y)$,
due to the fact that the rational invariants $\bOm_{\rm S}(\alpha_i,y)$ are rational functions of $y$ when $\alpha_i$ is not primitive. It is possible to give a minimal prescription that cures this problem by introducing new `scaling' contributions as follows \cite{Manschot:2011xc}
\be
\label{tentativeC2}
\begin{split}
\bar \Omega(\gamma;y) &= \sum_{n\geq 1} 
\sum_{\sum \alpha_i=\gamma}
\frac{\gref(\{\alpha_i\},\{c_i\},y)}{|{\rm Aut}(\{\alpha_i\})|} \, \\
\times  \prod_{i=1}^n  &
\left\{  \sum_{m_i|\alpha_i}  \frac{y-1/y}{m_i(y^{m_i}-y^{-m_i})}
 \left[ \Omega_{\rm S}(\alpha_i/m_i;y^{m_i}) 
 \empha{+ \Omega_{\rm scaling}(\alpha_i/m_i;y^{m_i})} \right]
 \right\}\ .
 \end{split} 
\ee
The new contribution $\Omega_{\rm scaling}$ is determined recursively in terms of $\Omega_{\rm S}$ by
\be \label{essp2}
\Omega_{\rm scaling}(\alpha;y) =
\sum_{\{\beta_i\in \Gamma\}, \{m_i\in\IZ\}\atop
m_i\ge 1, \, \sum_i m_i\beta_i =\alpha}
\emphb{H(\{\beta_i\}; \{m_i\};y)} \, \prod_i 
\Omega_{\rm S}(\beta_i;y^{m_i})
\ee
where the rational function
$H(\{\beta_i\}; \{m_i\};y)$ is in turn determined uniquely by the `minimal modification hypothesis',  
\begin{itemize}
\item[i)] $H$ is symmetric under $y\to 1/y$ and vanishes as $y\to 0$,
\item[ii)] the coefficient of $\prod_i \Omega^{\rm S}_{\rm ref}(\beta_i;y^{m_i})$ in the
expression for $\Omega(\sum_i m_i\beta_i; y)$ obtained using \eqref{tentativeC2} must be a 
Laurent polynomial in $y$.
\end{itemize}
It can be checked that the condition ii) exactly fixes the free parameters allowed by i). As an example,
for the 3-center configuration discussed in \S\ref{sec_scal}, 
\be 
\label{H3min}
\begin{split}
H(\{\gamma_1,\gamma_2,\gamma_3\}; &\{1,1,1\}; y) =  
\begin{cases} - 2   \,  (y-y^{-1})^{-2}\, \ , \hbox{$a+b+c$ even}\cr
(y + y^{-1}) \, (y-y^{-1})^{-2} \, \ , \hbox{$a+b+c$ odd}
\end{cases}
\end{split}
\ee
so the prescription r\eqref{tentativeC2} reduces to the replacement $\gref\to \tilde g_{\rm Coulomg}$, 
but this is not so in general. The Mathematica package developed in \cite{Manschot:2013sya} 
provides a convenient implementation of this prescription.

The Coulomb branch formula \eqref{tentativeC2} conjecturally expresses the set of indices $\Omega(\gamma;t,y)$ in terms of a new set of indices $\OmS(\gamma;y)$. One might ask, what have we gained ? First, unlike the total index $\Omega(\gamma;t;y)$, the single-centered invariants $\OmS(\gamma;y)$ no longer depend on the moduli. 
This may be a valuable property in enforcing invariance under various dualities, since these dualities typically act both on the charge $\gamma$ and moduli $t$. Second, in $\cN=2$ string vacua we expect that 
$\OmS(\gamma;y)$ counts micro-states of single-centered BPS black holes, with $AdS_2$ near-horizon geometry. Supersymmetry requires these BPS microstates to lie in singlets of the rotation group \cite{Dabholkar:2010rm}, thus
their refined index $\OmS(\gamma;y)$ should be independent of $y$. If this hypothesis is correct, then 
the single-centered invariants $\OmS(\gamma)$ provide a concise parametrization of the BPS spectrum in all chambers of moduli space, which can be directly compared with the path integral of quantum gravity in a single $AdS_2$ throat, also known as the quantum entropy function \cite{Sen:2008yk}. 
In the next lecture, we shall test this
hypothesis in the case of quiver quantum mechanics, which  provides a model for  D-brane bound states.

\section{Applications to quivers}

In this last lecture, we shall apply the Coulomb branch formula \eqref{tentativeC2} in the context of 
quiver quantum mechanics (or, in mathematical parlance, the cohomology of quiver moduli spaces), where
the refined index $\Om(\gamma;t;y)$ is unambiguously defined and often computable. Quiver quantum
mechanics on the Higgs branch provides an accurate description of D-brane bound states near loci in moduli space where the central charges of the constituents nearly align \cite{Denef:2002ru}. On the Coulomb branch, it also reproduces the quantum mechanics of multi-centered solitons discussed in \S\ref{sec_qm}. This physical duality has very interesting mathematical consequences for the  cohomology of quiver moduli spaces, as we
shall explain.

\subsection{Quiver quantum mechanics}

Quiver quantum mechanics is a particular  type of matrix quantum mechanics with $\cN=4$ supercharges,
 whose matter content is 
encoded in a quiver, \footnote{Quivers also describe SUSY gauge theories in higher dimension, but here we focus on $D=1$}, i.e. a set of nodes and arrows (see e.g. \eqref{KronQuiv} or \eqref{xy3node} below).  
Each node $\ell=1...K$ represents a $U(N_\ell)$ vector multiplet $(\vec r_\ell, D_\ell)$, each arrow 
represents a chiral multiplet $\phi_{\kk,\ell}$ in $(N_\ell, \bar N_\kk)$ representation of $U(N_\ell)\times U(N_\kk)$. 
The set $\{N_\ell\}$ is called the dimension vector.
In addition, one must specify \emphb{Fayet-Iliopoulos terms} $c_\ell$ such that $\sum_\ell N_\ell c_\ell=0$, and (in presence of closed oriented loops) a gauge invariant \emphb{superpotential} $W(\phi_{\kk,\ell})$. 
In applications to D-brane
bound states, each node represents $N_\ell$ coinciding D-branes of charge $\gamma_\ell$ and the number
of arrows from node $\ell$ to $\kk$ is given by the DSZ product\footnote{In general, the DSZ product  $\gamma_{\ell\kk}$ gives the {\it net} number of arrows from node $\ell$ to $\kk$, counted with orientation,
but we shall always assume that the superpotential $W$ is generic such that chiral multiplets associated to pairs of arrows with opposite
direction are massive and can be integrated out.}
  $\gamma_{\ell\kk}$. 
The FI term $c_\ell$ depends on  vector multiplet (VM) moduli, while the coefficients of $W$
in general depend on hypermultiplet (HM) moduli. 

On the Coulomb branch,  the chiral multiplets and off-diagonal components of the vector multiplets 
are massive and can be integrated out. The diagonal components of the vector multiplets
$\vec r_\ell$ then represent the locations of $N_\ell$ Abelian D-branes carrying charge $\gamma_\ell$, subject
to the D-term constraints 
\be 
\forall \ell\ ,\quad \sum_{\kk\ne \ell} \frac{\gamma_{\ell\kk}}{|\vec r_\ell -\vec r_\kk|}
=  c_\ell\ .
\ee
Remarkably, this reproduces the equations \eqref{denef3d} satisfied by multi-centered black hole solutions of $\cN=2$ supergravity  \cite{Denef:2002ru}. This Coulomb branch description is valid provided the centers
are kept far apart.

The Higgs branch description, on the other hand, is valid for large values of the chiral multiplet scalars $\phi_{\kk\ell}$, such that  the vector multiplets can be integrated out. 
The moduli space of SUSY vacua on the Higgs branch $\cM_{H}$ is the set of solutions of the F-term \emphc{$\pa_{\phi} W=0$} and D-term equations 
\be \label{emodi1}
\begin{split}
\forall \ell:\ 
\sum_{ \gamma_{\ell\kk}>0} \phi_{\ell\kk}^* \, T^a \, 
\phi_{\ell\kk} - \sum_{ \gamma_{\kk\ell}>0} 
\phi_{\kk\ell}^* 
\, T^a \, 
\phi_{\kk\ell,\alpha,s't} = c_\ell \, \Tr(T^a)
\end{split}
\ee
modulo the action of $\prod_\ell U(N_\ell)$. 
Equivalently, $\cM_H$ is the space of semi-stable solutions of $\pa_\phi W=0$ modulo
 $\prod_\ell GL(N_\ell,\IC)$.
BPS states on the Higgs branch correspond to \empha{cohomology classes} 
in $H^*(\cM_{\rm H},\IZ)$.
The refined index on the Higgs branch is given by the Poincar\'e  polynomial
(more accurately, a polynomial in $y,1/y$, symmetric under $y\to 1/y$)
\be
\label{defQ}
Q(\cM_{H};y) = \Tr'(-y)^{2J_3} = \sum_{p=1}^{2d} b_p(\cM_H) \, (-y)^{p-d}\ .
\ee
Here $d$ is the complex dimension of $\cM_H$, given  by
\be
d = \sum_{\gamma_{\ell\kk}>0} \gamma_{\ell \kk}\, N_\ell N_\kk - \sum_{\ell} N_l^2 - f +1\ ,
\ee
where $f$ is the number of independent F-term conditions ($f=0$ for quivers without loop).
Since $\cM_H$ is a K\"ahler manifold, the total cohomology $H^*(\cM_{\rm H},\IZ)$ admits an 
action of  $SU(2)$ known as the \emphb{Lefschetz action}, 
\be
J_+\cdot h=\omega_K \wedge h\ , \quad 
J_- = \omega_K \,\llcorner\, h\ ,\quad J_3\cdot h = \tfrac12 (n-d) h\ .
\ee
where $\omega_K$ is the K\"ahler form on $\cM_{H}$ and $n$ is the form degree of $h$.
The parameter $y$ in \eqref{defQ} is conjugate to $2J_3$, and therefore $Q(\cM_{H};y)$
must decompose as a sum of characters of spin $j$ representations of $SU(2)$ with $j\leq d/2$.

To demystify these definitions,
consider the simplest example of the 2-node  quiver with $k$ arrows (also known as the
generalized Kronecker quiver), with dimension vector (1,1):
\be
\label{KronQuiv}
\begin{xy} 0;<1pt,0pt>:<0pt,-1pt>:: 
(0,0) *+{1} ="0",
(98,0) *+{1} ="1",
"0", {\ar|*+{\scriptstyle k}"1"},
\end{xy}
\ee
The vacuum moduli space $\cM_{H}$ is the hypersurface $\sum_{\ell=1}^k |\phi_\ell|^2=c_1$ in $\IC^k$, modded out by $U(1)$ rotations $\phi_\ell\mapsto e^{\I\theta} \phi_\ell$. For $c_1<0$, $\cM_{H}$ is empty, whereas for 
$c_1>0$, $\cM_{H}= \IC^k/\IC^\times = \IP^{k-1}$. The complex projective space $\IP^{k-1}$
has $b_{i}=1$ for $i$ even, $0\leq i\leq 2(k-1)$ and $b_i=0$ for $i$ odd. Its Poincar\'e polynomial is therefore
\be
\label{QKron}
Q(\cM_{H};y) = \frac{ (-y)^{k} - (-y)^{-k}}{y-1/y} \ ,
\ee
which is recognized as the character of a spin $j=(k-1)/2$ representation of the Lefschetz 
$SU(2)$ action. The fact that \eqref{QKron} agrees with the Coulomb index \eqref{gref2} for two 
centers is not a coincidence, as we shall see shortly. For $k=2$ and dimension vector $(M,N)$, 
one can show that $Q(\cM_{H};y)$ is the number of BPS states of charge $M\gamma_1+N\gamma_2$ 
in the weak coupling chamber of pure $SU(2)$ Seiberg-Witten theory, again not a coincidence !

In order to apply the Coulomb branch formula \eqref{tentativeC2} to quiver moduli spaces, we just need to
identify the Poincar\'e polynomial $Q(\cM_{H};y)$ with the total index $\Omega(\gamma;y)$, where the charge $\gamma$ is identified with the dimension vector $\{N_\ell\}$ in the positive cone  $\Gamma=\IN^K$ of the
charge lattice $\IZ^K$. Eq. \eqref{tentativeC2} then predicts that the Poincar\'e polynomial $Q(\cM_{H};y)$
should be expressible in terms of `single centered invariants' $\OmS(\alpha)$, 
associated to quivers with the same topology but with smaller dimension vectors $\alpha=\{M_\ell\}$, 
with $M_\ell\leq N_\ell$ for all $\ell=1,\dots K$. While the total indices $\Omega(\gamma;y)$ can always
be traded for the `single centered invariants' $\OmS(\alpha)$ using \eqref{tentativeC2}, it is highly non-trivial
that the $\OmS(\alpha)$'s are independent both of the FI parameters $c_\ell$ and of the Lefschetz parameter
$y$. In the rest of this lecture we shall test this conjecture on a variety of quivers for which $Q(\cM_{H};y)$
is computable.

\subsection{Quivers without loops}

For quivers without loops, and arbitrary  dimension vector, the Poincar\'e polynomial can be computed 
recursively using Harder-Narasimhan filtrations, and counting of points over finite fields \cite{1043.17010}. 
This sophisticated
mathematics leads to a very explicit formula which can be stated as follows  \cite{Manschot:2013sya}:
\be
\label{eq:inversestackinv}
\begin{split}
\bar Q\left(\{ N_\ell \} ; \{ c_\ell \} ;y\right) 
  =& \sum_m
  \sum_{\{\vec M^{(i)}\} \atop
  \sum_{i=1}^m \vec M^{(i)}=\vec N; \vec M^{(i)}\parallel \vec N
  \,\mathrm{for}\,\, i=1,\dots,m}
\frac{1}{m\,(y-1/y)^{m-1}}  \prod_{i=1}^m 
\gR(\{ \{ M^{(i)}_\ell \};  \{c_\ell\};y)
\, ,
\end{split}
\ee
where the sum runs over all {\it ordered} partitions of
 $\vec N\equiv (N_1,\cdots N_K)$ into parallel vectors $\vec M^{(1)}, \cdots \vec M^{(\ell)}$. On the r.h.s,
  $\gR(\{ \{ M^{(i)}_\ell \};  \{c_\ell\};y)$ is the `stack invariant' associated to the quiver 
with dimension vector $\vec M^{(i)}$, given by 
  \be \label{ec5}
\begin{split}
 \gR& (\{M_\ell\}; \{c_\ell \};y) 
=  (-y)^{-  \sum_{i,j} M_i M_j \max(\gamma_{ij},0) -1 +\sum_i M_i} \,
\\  & \times  (y^2-1)^{1-\sum_i M_i} \, \sum_{\rm partitions}  (-1)^{s-1} y^{2\sum_{a\leq b}\sum_{i,j}
\max(\gamma_{ij},0) \, 
N^b_i N^a_j} \prod_{a,i} ([N^a_i,y]!)^{-1}\, .
\end{split}
\end{equation}
Here the sum runs over all {\it ordered} partitions of 
the vector $(M_1,\cdots M_L)$ into 
non-zero vectors $\{(N^a_1,\cdots N^a_K), \quad a=1,\dots, s\}$
for $s=1,\dots, \sum_i {M_i}$,
satisfying $N^a_i\ge 0$, 
$\sum_a N^a_i=M_i$ and \be \label{econdombetana} 
\sum_{a=1}^b \, \sum_{i=1}^K N^a_i  c_i > 0 
\ee 
for all $b$ between 1 and $s-1$.
$[N,y]!$ denotes the $q$-deformed factorial,
\be \label{ec5.5}
[N,y]! \equiv
[1,y][2,y]\ldots[N,y]\, ,\qquad 
[N,y] \equiv \frac{y^{2N}-1}{y^2-1}\ .
\ee
When the charge vector $\{N_\ell\}$ is primitive, $\bar Q\left(\{ N_\ell \} ; \{ c_\ell \} ;y\right)$ coincides with
$\gR (\{N_\ell\}; \{c_\ell \};y)$ and the corresponding $Q\left(\{ N_\ell \} ; \{ c_\ell \} ;y\right)$ is indeed a symmetric
polynomial in $y,1/y$ with alternating sign integer coefficients.  If $\{N_\ell\}$ is not primitive, then the
quiver moduli space $\cM_{\rm H}$ is singular, and it is more tricky to define its cohomology, nevertheless
\eqref{eq:inversestackinv}  produces a bona-fide Poincar\'e polynomial in all cases. The Mathematica
package supplied with \cite{Manschot:2013sya} gives a convenient implementation of the
`Higgs branch formula' \eqref{eq:inversestackinv}.

In order to compare the exact  answer \eqref{eq:inversestackinv} with the prediction from the Coulomb
branch formula \eqref{tentativeC2}, a crucial fact is that the stack invariants \eqref{ec5} with arbitrary
dimensions $\{M_\ell\}$ can all be computed in terms of stack invariants of Abelian quivers, a property
known as the Abelianization (or MPS) formula  \cite{Manschot:2010qz,ReinekeMPS} 
\be \label{ehiggsexp}
\gR(\{N_i\}; \{\gamma_i\};\{\zeta_i\};y) =\!\!\!\!\!\!\!\!
\sum_{\{k^{(\ell)}_j\}\atop \sum_\ell \ell k^{(\ell)}_j=N_j}\!\!\!\!\!\!\!\!
\gR(\{1\}, \{ (\ell\gamma_j)^{k^{(\ell)}_j}\}; \{ (\ell \zeta_j)^{k^{(\ell)}_j}\}; y)
\prod_{i=1}^K 
\prod_\ell {1\over k^{(\ell)}_i!}\left( {y - y^{-1}\over \ell (y^\ell - y^{-\ell})}\right)^{k^{(\ell)}_i}\, .
\ee
The quivers appearing on the r.h.s. are obtained by applying the following prescription on each node of the
original quiver:
\begin{itemize}
\item Each vertex $i$ with dimension $N_i$  is replaced by a
collection $i_{\ell,k}$ of vertices with $k=1,\dots k_\ell$ for any $\ell$ in the partition 
$N_i=\sum_\ell \ell\, k_{\ell}$
\item Each  arrow $i\to j$  (resp. $j\to i$) with $j\neq i$ in the original quiver induces $\ell$ arrows $i_{\ell,k}\to j$ (resp. $j\to i_{\ell,k}$) in the new quiver for every $\ell,k$;
\item The FI parameter at the nodes $i_{\ell,k}$  is equal to $\ell c_i$;
\end{itemize}
Using this formula, one can prove that the Higgs branch formula \eqref{eq:inversestackinv} agrees with the Coulomb branch formula \eqref{tentativeC2} provided the only non-vanishing single-centered invariants $\OmS(\gamma)$ are those associated to the basis vectors $\gamma_\ell$, with dimension 1
at node $\ell$ and dimension 0 at the other nodes. These single-centered invariants are moreover
moduli-independent and carry zero angular momentum, namely
\be
\OmS(\gamma_\ell;y) = 1 \ ,\quad \OmS(\gamma)=0 \quad \mbox{if} \quad \gamma\notin \{\gamma_\ell, \ell=1\dots K\}\ .
\ee
Thus, the complete BPS spectrum on the Higgs branch of quiver quantum mechanics associated to quiver without loops can be understood from the Coulomb branch picture, where all the states are bound states of single-centered black holes carrying unit BPS index. For Abelian quivers, this equivalence is expected since  the Coulomb branch and the Higgs branch are both regular and disconnected, and have overlapping regime of 
validity \cite{Denef:2002ru}. The simplest case of this equivalence is the Abelian
Kronecker quiver \eqref{KronQuiv}, with Higgs branch $\IP^{k-1}$, whose de Rham cohomology 
is isomorphic to that the Dirac cohomology of $S^2$ with $k$ units of flux. More generally,
one can show that the stack invariant \eqref{ec5} for Abelian quivers agrees with the Coulomb index
\eqref{eindexfin}. 
The fact that this equivalence
holds also for non-Abelian quivers is more striking, as the Higgs branch and Coulomb branch are no longer
disconnected.

\subsection{Quivers with loops}

For general quivers with closed oriented loops and generic superpotential, there is (to my knowledge) no general way of computing the cohomology of the quiver moduli space.  For Abelian quivers however, the situation is much simpler, since the quiver moduli space $\cM_H$ arises as a complete intersection (corresponding to the F-term
constraints) in a product of projective spaces $\cM_{\rm amb}$
(representing the solution of the D-term constraints modulo
gauge symmetries). In such cases, the Poincar\'e polynomial can be computed by combining the Lefschetz
hyperplane theorem, which determines all Betti numbers except the middle one,
\be
\label{hypthm}
b_p(\cM_H) = \begin{cases} b_p(\cM_{\rm amb}) & p<d \\
b_{2d-p} (\cM_{\rm amb}) & p>d\ ,
\end{cases}
\ee
and the Riemann-Roch theorem to compute $\chi(\cM_{H})$ and hence the middle cohomology
$b_{d}(\cM_H)$. Using these techniques, the cohomology of the simplest Abelian quiver with loop
\be
{\scriptsize \label{xy3node}
\begin{xy} 0;<.5pt,0pt>:<0pt,-.5pt>:: 
(38,0) *+{1} ="0",
(79,67) *+{1} ="1",
(0,69) *+{1} ="2",
"0", {\ar|*+{a}"1"},
"2", {\ar|*+{c}"0"},
"1", {\ar|*+{b}"2"},
\end{xy}} 
\ee
was studied in \cite{Denef:2007vg} and revisited in \cite{Bena:2012hf}. Denoting by
$\cM_{a,b,c}$ the quiver moduli space in the chamber $c_1,c_2>0, c_3<0$, it was found that
the generating function of the Poincar\'e polynomials
\be
Q(x_1,x_2,x_3;y) = \sum_{a\geq 0,b\geq 0,c\geq 0} (x_1)^a (x_2)^b (x_3)^c 
\, Q(\cM_{a,b,c},y) 
\ee
decomposes into a sum $Q=Q_{\rm C} + Q_{\rm S}$, where
\be
\begin{split}
Q_{\rm C}(x_1,x_2,x_3;y) =& \frac{x_1 x_2\left\{1-x_1x_2+x_1 x_2 x_3 \left(x_1+x_2 +
    y + y^{-1}
    \right)\right\}}{(1-x_1 x_2) (1-x_1 x_3)(1-x_2 x_3) (1+x_1/y)
    (1+x_1 y) (1+x_2/y) (1+x_2 y)}\ , \\
Q_{\rm S}(x_1,x_2,x_3;y)=&\frac{x_1^2 x_2^2 x_3^2}{
(1-x_1 x_2)(1-x_2 x_3)(1-x_1 x_3)[1-x_1x_2-x_2 x_3-x_1 x_3- 2x_1x_2x_3]}\ .
\end{split}
\ee
The first term $Q_{\rm C}$ can be understood as the partition function of the Coulomb indices
\eqref{elocal55} of 3-centered black holes carrying charges $\gamma_1,\gamma_2,\gamma_3$ 
corresponding to the nodes of the quiver, with unit degeneracy $\OmS(\gamma_i;y)=1$. 
This contribution is both $y$ dependent and moduli dependent (as manifest from the lack of invariance
under circular permutations of $x_1,x_2,x_3$), and its Taylor coefficients grow polynomially with $a,b,c$. 
The second term $Q_{\rm S}$, on the other hand, is $y$-independent, therefore associated to 
Lefschetz singlets in the middle cohomology of $\cM_{a,b,c}$. Remarkably, it is  invariant under circular permutations of $x_1,x_2,x_3$, which shows that  the`pure-Higgs' states are 
robust under wall-crossing. They can be shown to contribute only when the triangular inequalities $a<b+c,b<a+c,c<a+b$ are satisfied, that is the region where scaling solutions occur. Moreover, 
their contribution to $Q(\cM_{a,b,c},y)$  grows  exponentially as $2^{a+b+c}$ under overall rescaling of $a,b,c$. The total result $Q=Q_{\rm C} + Q_{\rm S}$ agrees precisely with the prediction of the 
Coulomb branch formula \eqref{tentativeC2}, 
\be \label{efour}
Q(\gamma_1+\gamma_2+\gamma_3; y) 
= \gref (\gamma_1, \gamma_2, \gamma_3; y)  + H(\{\gamma_1,\gamma_2,\gamma_3\};\{1,1,1\}; y)
+ \OmS(\gamma_1+\gamma_2+\gamma_3)
\ee
provided $H$ is chosen according to the minimal modification hypothesis \eqref{H3min}, and the single-centered index $\OmS(\gamma_1+\gamma_2+\gamma_3)$ is equated with the number of  Lefschetz singlets. This study
was extended to all Abelian `necklace' quivers in \cite{Lee:2012sc,Lee:2012naa,Manschot:2012rx}
with similar results.

In \cite{Manschot:2012rx} we have tested the formula on a variety of Abelian quivers with loops, as well as on some non-Abelian quivers, and we have found that the Coulomb branch formula \eqref{tentativeC2} correctly
predicts the cohomology of the quiver moduli space, with constant, $y$-independent elementary or
`pure-Higgs' invariants $\OmS(\gamma)$ associated to scaling subquivers. It is worth stressing that unlike the case of necklace quivers, the invariants $\OmS(\gamma)$ in general do not only contribute to the middle cohomology, as they occur as coefficients of  non-trivial $SU(2)$ characters in \eqref{tentativeC2}.

\section{Conclusion and open problems}

In these lectures, we have seen that the simple physical picture of BPS states as bound states of elementary constituents carrying their own degrees of freedom and interacting via the supersymmetric quantum mechanics
of charged particles is remarkably powerful: it provides a simple derivation of the wall-crossing formula, and allows to parametrize the BPS index $\Omega(\gamma;t)$ for all values of the parameters or moduli in terms of certain chamber-independent, elementary degeneracies $\OmS(\gamma)$. Similarly  the protected spin character in $\cN=2$ gauge theories, or the refined index $\Omega(\gamma;t;y)$ in 
$\cN=2$ string vacua, can be parametrized in terms of refined elementary degeneracies $\OmS(\gamma;y)$.
While the refined index can get contributions both from short and long multiplets, it seems plausible that it should satisfy the same wall-crossing properties as the protected spin character as a function of the vector multiplet moduli. On the other hand, it is quite possible that  the elementary degeneracies $\OmS(\gamma;y)$ may jump on certain loci in hypermultiplet moduli space, in particular in interpolating from small to large string coupling
(at weak coupling, these jumps can only occur in co-dimension 2, since the  hypermultiplet moduli only
affect the holomorphic superpotential $W$). 
The large string coupling regime is the one where a gravitational picture may apply, and where a comparison
with the quantum entropy function may be carried out. Unfortunately,  the Coulomb branch formula \eqref{tentativeC2}  remains rather impractical, as it requires an enumeration of all possible bound states
with regular metric. 

In the case of BPS states described at weak coupling by quiver quantum mechanics, the Coulomb branch formula 
is completely well-defined, albeit still conjectural. In all examples that we have analyzed, it is satisfying to see that the the elementary degeneracies $\OmS(\gamma;y)$ computed from the Coulomb branch formula are  independent of $y$, in agreement with the large coupling prediction that single-centered BPS black holes have zero angular momentum. From a mathematical point of view, this is a very unexpected property
of quiver moduli spaces, whose understanding may help unravel new algebraic structures. From a physics point of view, it would be very interesting to clarify the relation between the Higgs and Coulomb branch descriptions
and derive the Coulomb branch formula \eqref{tentativeC2} from quiver quantum mechanics, as it would
provide a solvable example of open/closed string duality. 

On a longer time-scale, it will be exciting to apply these techniques to actual $\cN=2$ string vacua, such as type II 
strings compactified on Calabi-Yau threefold, and test agreement between macroscopic and microscopic descriptions of BPS black holes, to the same level of precision which was achieved for $\cN\geq 4$ string 
vacua, and possibly address non-BPS black holes as well.

\acknowledgments

I am greatly indebted to Jan Manschot and Ashoke Sen for a very enjoyable collaboration on the results reviewed herein. I am also grateful to the organizers of the Corfu Summer Institute for the kind invitation to lecture and excellent hospitality.


\providecommand{\href}[2]{#2}\begingroup\raggedright\endgroup

\end{document}